%% file: arc24.tex
\documentclass[runningheads]{llncs}
\usepackage[T1]{fontenc}
\usepackage{graphicx}
\usepackage{amsmath}
\usepackage{algorithmic}
\usepackage{textcomp}
\usepackage{booktabs}
\usepackage{tabularx}
\usepackage{array}
\usepackage{threeparttable}
\usepackage{subfig} 
\usepackage{xcolor}
\usepackage{url}
\usepackage{hyperref}
\usepackage{scalerel}
\usepackage{tikz}
\usetikzlibrary{shapes.geometric, arrows} 
\usepackage{float}
\usepackage{placeins}
\usepackage{comment}
\usepackage{multirow}
\usepackage{tabu}
\usepackage{colortbl}
\usepackage[linesnumbered,ruled,noend]{algorithm2e}
\usepackage{pifont}
\usepackage{IEEEtrantools}
\usepackage{cite}

\newcommand{\thickhline}{%
    \noalign {\ifnum 0=`}\fi \hrule height 1pt
    \futurelet \reserved@a \@xhline
}

\SetCommentSty{mycommfont}    
{}

\title{NEUROSEC: FPGA-Based Neuromorphic Audio Security}
\author{
Murat Isik\inst{1} \and
Hiruna Vishwamith\inst{2} \and
Yusuf Sur\inst{3} \and
Kayode Inadagbo\inst{4} \and
I. Can Dikmen\inst{5}
}
\authorrunning{Isik et al.}
\institute{
Drexel University, Philadelphia, PA, USA \\
\email{mci38@drexel.edu} \and
University Of Moratuwa, Moratuwa, Sri Lanka \\
\email{vishwamithpgh.20@uom.lk} \and
Abdullah Gul University, Kayseri, Turkey \\
\email{yusuf.sur@agu.edu.tr} \and
Prairie View A\&M University, Prairie View, TX, USA \\
\email{kayodeinadagbo@gmail.com} \and
Temsa R\&D Center, Adana, Turkey \\
\email{can.dikmen@temsa.com}
}

\begin{document}
\maketitle

\begin{abstract}
    \input{arc2024/sections/abstract}
\end{abstract}


\section{Introduction}
\input{arc2024/sections/introduction}

\section{Neuromorphic Hardware: Evolution, Applications, and Security}
\label{sec:nhw}
\input{arc2024/sections/nhw}

\subsection{Security Challenges}
\label{sec:security}
\input{arc2024/sections/security}

\section{Proposed Design Methodology}
\label{sec:design_methodology}
\input{arc2024/sections/design_methodology}

\section{Evaluation}
\label{sec:evaluation}
\input{arc2024/sections/evaluation}

\section{Conclusions}
\label{sec:conclusions}
\input{arc2024/sections/conclusions}

\bibliographystyle{splncs04}
\bibliography{external}

\end{document}

%% file: arc2024/sections/abstract.tex
Neuromorphic systems, inspired by the complexity and functionality of the human brain, have gained interest in academic and industrial attention due to their unparalleled potential across a wide range of applications. While their capabilities herald innovation, it is imperative to underscore that these computational paradigms, analogous to their traditional counterparts, are not impervious to security threats. Although the exploration of neuromorphic methodologies for image and video processing has been rigorously pursued, the realm of neuromorphic audio processing remains in its early stages. Our results highlight the robustness and precision of our FPGA-based neuromorphic system. Specifically, our system showcases a commendable balance between desired signal and background noise, efficient spike rate encoding, and unparalleled resilience against adversarial attacks such as FGSM and PGD. A standout feature of our framework is its detection rate of 94\%, which, when compared to other methodologies, underscores its greater capability in identifying and mitigating threats within 5.39 dB, a commendable SNR ratio.  Furthermore, neuromorphic computing and hardware security serve many sensor domains in mission-critical and privacy-preserving applications.

%% file: arc2024/sections/introduction.tex
Computer hardware that emulates the intricate functions of the human brain has been termed neuromorphic hardware. Drawing inspiration from biological neural systems, neuromorphic hardware aims to replicate the way these systems process information, bridging the gap between biological cognition and artificial computation. Neuromorphic computing represents a paradigm shift from traditional computing methodologies. At its core, it seeks to emulate the brain's neural structures and functionalities, offering a more natural and efficient way to process information. The significance of neuromorphic computing lies in its potential to revolutionize various domains, from artificial intelligence to robotics, by providing systems that can learn, adapt, and evolve in real-time \cite{huynh2022implementing, marchisio2021dvs, inadagbo2023exploiting}. Field-Programmable Gate Arrays (FPGAs) have emerged as a pivotal component in the neuromorphic computing landscape. Their inherent reconfigurability and parallel processing capabilities align seamlessly with the demands of neuromorphic systems. FPGAs offer the flexibility to design and customize neuromorphic architectures, enabling researchers and engineers to experiment with and optimize neural network designs, thereby pushing the boundaries of what neuromorphic systems can achieve. As with any computing system, security remains paramount in neuromorphic systems. Given their potential applications in sensitive areas such as defense, healthcare, and finance, ensuring the integrity, confidentiality, and availability of data processed by neuromorphic systems is crucial. Furthermore, the unique architecture and operation of neuromorphic systems present both challenges and opportunities in the realm of security, requiring specialized approaches to safeguard them against threats \cite{salehi2022neuromorphic, merchant2022security, chen2021adversarial, 10268746, isik2023astrocyte, isik2022design}. Neuromorphic hardware, inspired by the intricate functions of the human brain, seeks to bridge the gap between biological cognition and artificial computation. This approach represents a paradigm shift, offering a more natural and efficient way to process information. FPGAs, with their inherent reconfigurability and parallel processing capabilities, have emerged as a pivotal component in this landscape, enabling the design and customization of neuromorphic architectures. Given the potential applications of neuromorphic systems in sensitive areas such as defense and healthcare, ensuring their security is paramount. The unique architecture of these systems presents both challenges and opportunities in the realm of security.

\vspace{5pt}\textbf{In this paper, we present the following contributions:}

\begin{itemize}
\item We explore SNN-based neuromorphic audio processing, a niche compared to image/video processing.
\item We analyze security threats in neuromorphic audio, emphasizing adversarial attacks like FGSM and PGD that introduce audio artifacts.
\item Our FPGA-integrated system boasts a 94\% detection rate, efficient spike encoding, and a balanced signal-to-noise ratio.
\item We compare our framework with existing methods, highlighting its superior threat detection and mitigation within a favorable SNR.
\end{itemize}

%% file: arc2024/sections/nhw.tex
Neuromorphic hardware has transitioned from basic silicon neurons to sophisticated neuromorphic chips, offering benefits, especially in security. FPGAs enhance these systems with their flexibility in design and parallel processing capabilities. Several studies have explored the integration of neuromorphic systems with FPGAs, touching upon design methodologies, applications, and security implications. The journey of neuromorphic computing began with the vision of replicating the brain's neural structures in silicon. Early endeavors focused on creating silicon neurons, aiming to capture the parallel processing capabilities of the brain. Over time, advancements in technology and research led to the development of advanced neuromorphic chips, which are now at the forefront of many cutting-edge applications. FPGAs offer flexibility in design, allowing for the customization of neuromorphic architectures. Their reconfigurability and parallel processing capabilities align well with the inherent characteristics of neuromorphic systems. Several studies and research endeavors have delved into the integration of neuromorphic systems with FPGAs. These works have explored various aspects, from design methodologies to applications, and have also touched upon the security implications of such integrations. Neuromorphic hardware offers a range of benefits, especially in the context of security. Neuromorphic hardware, with its unique architecture and capabilities, holds immense promise in the realm of security. Its integration with FPGAs further amplifies its potential, paving the way for adaptive security solutions \cite{peng2023pasnet, gongye2023hammerdodger, zhou2023nnsplitter, isik2023survey}. Researchers explored the use of temporal dependency in audio data to mitigate the impact of adversarial examples, particularly in automatic speech recognition (ASR) systems. The study shows that input transformations, often used in image adversarial defense, provide limited robustness improvement in audio data and are susceptible to advanced attacks. Conversely, exploiting temporal dependencies in audio can effectively discriminate against adversarial examples and resist adaptive attacks on Recurrent Neural Network (RNN) \cite{yang2018characterizing}. It was showed the vulnerability of Deep Neural Networks (DNNs) to adversarial examples, particularly in the audio field. Adversarial examples are crafted by adding subtle noise to original samples, which can deceive machines while remaining imperceptible to humans. The paper proposes a defense method that introduces low-level distortion via audio modification to detect these adversarial examples. The idea is that while the classification of the original sample remains stable under this distortion, the adversarial example's classification changes significantly. This method was tested using the Mozilla Common Voice dataset and the DeepSpeech model, showing a significant drop in the accuracy of adversarial examples, thereby effectively detecting them \cite{kwon2019poster}. U-Net based attention model were introduced for enhancing adversarial speech signals. The proposed self-attention speech U-Net is designed to improve the robustness against adversarial examples in speech recognition systems. The model uses attention mechanisms in its upsampling blocks to better process adversarial noise in speech signals. The study demonstrates that while traditional methods of speech enhancement can increase signal-to-noise ratio (SNR) scores, they often fail to improve other key metrics such as PESQ, STI, and STOI. The authors also found that adversarial training can further enhance the performance of the Convolutional Neural Network (CNN), making it more robust against adversarial attacks in speech recognition \cite{yang2020characterizing}.

\begin{table}[h]
\caption{Features of Neuromorphic Systems.}
\centering
\small 
\begin{tabular}{|l|l|} 
\hline
\textbf{Feature} & \textbf{Description} \\
\hline
Speed & Emulates brain for fast parallel processing. \\
\hline
Power Efficiency & Energy-efficient chips for continuous monitoring. \\
\hline
Adaptive Learning & Evolves algorithms for new threats. \\
\hline
Anomaly Detection & Flags deviations as threats. \\
\hline
Hardware Security & Robust protection via FPGA integration. \\
\hline
Parallel Processing & Processes multiple data streams. \\
\hline
Scalability & Supports expansion for security needs. \\
\hline
Resilience & Resists conventional system attacks. \\
\hline
Real-time Response & Instant threat response. \\
\hline
Integration & FPGA versatility offers comprehensive security. \\
\hline
\end{tabular}

\label{table:neuromorphic_features}
\end{table}

\subsection{Spiking Neural Networks (SNNs)}

\begin{figure}[h!]
	\centering	\centerline{\includegraphics[width=0.99\columnwidth]{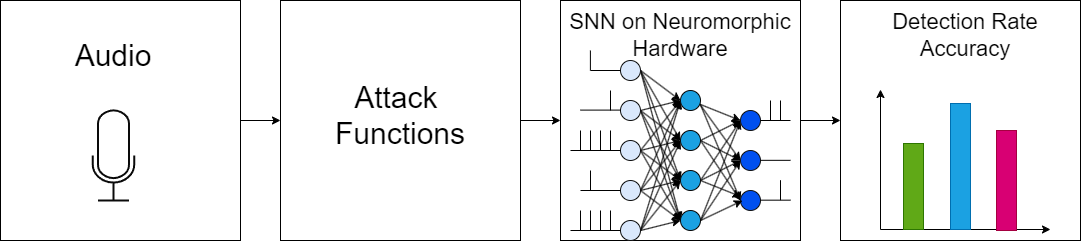}}
	\caption{Hardware Security Framework for Neuromorphic Audio Systems.}
	\label{fig:fig1}
\end{figure}

SNNs are designed to computationally emulate the behavior of biological neurons. As the intricacies of these networks grow, so do the computational demands associated with SNN inference. This growth has intensified the trade-off between hardware resources, power consumption, and acceleration performance, making it a focal point of contemporary research. Consequently, there's a burgeoning need for specialized hardware accelerators that can optimize computing-to-power efficiency ratios, especially in embedded and lightweight applications. One of the salient features of SNNs, from a hardware implementation perspective, is their communication mechanism. Neurons in SNNs communicate using spikes, which, in terms of logic resources, can be equated to a single bit, thereby reducing logic occupation. Recent studies have highlighted the potential of SNNs in enhancing security. For instance, researchers have shown that noise filters for Dynamic Vision Sensors (DVS) can act as defense mechanisms against adversarial attacks. They conducted experiments with various attacks, specifically in the setting of two different noise filters tailored for DVS cameras \cite{marchisio2021dvs}. In another notable study, a novel attack method tailored for rate coding SNNs was introduced, named the Rate Gradient Approximation Attack (RGA). This method was employed to detect abnormal traffic patterns, indicative of attacks, in Networks-on-Chip data using SNNs \cite{bu2023rate, madden2018adding}. Fig. \ref{fig:fig1} illustrates a generic framework for implementing hardware security in neuromorphic audio systems.

\subsection{Event-based Applications in Audio Processing}

\begin{figure}[h!]
	\centering
	\centerline{\includegraphics[width=0.9\columnwidth]{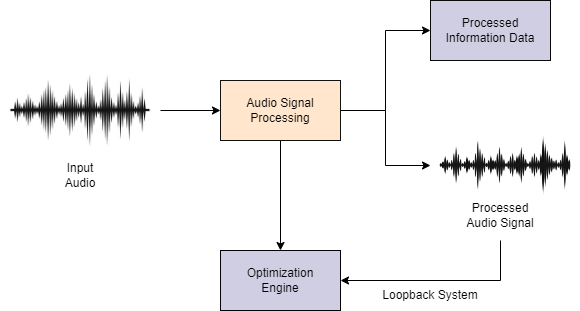}}
	\caption{Audio Processing Diagram.}
	\label{fig:fig2}
\end{figure}

Event-based audio processing is an emerging paradigm that draws inspiration from the asynchronous nature of the human auditory system which is depicted schematic representation of the audio processing workflow in Fig. \ref{fig:fig2}. Unlike traditional audio processing techniques that operate on uniformly sampled data, event-based methods focus on capturing and processing significant audio events as they occur. Researchers provide a comprehensive review of event-based sensing and signal processing across various sensory domains, including the auditory system \cite{tayarani2021event}. Their work explains the advantages of event-based approaches, especially in mimicking biological sensory systems, and offers insights into the potential applications and challenges of this paradigm. The authors delve into the post-processing of audio event detectors, employing reinforcement learning to enhance their performance \cite{giannakopoulos2022improving}. Their approach underscores the potential of combining advanced machine-learning techniques with event-based audio processing to achieve superior detection accuracy and efficiency. Furthermore, the significance of neuromorphic auditory computing in the context of robotics is highlighted in \cite{galan2022neuromorphic}. The author emphasizes the potential of a digital, event-based implementation of the hearing sense, paving the way for more responsive and adaptive robotic systems that can interact seamlessly with their environment.

%% file: arc2024/sections/security.tex
Neuromorphic systems have garnered significant attention due to their potential in various applications, from robotics to artificial intelligence. However, like all computing systems, they are not immune to security threats. This section delves into the unique security challenges posed by FPGA-based neuromorphic systems, drawing from existing literature and current research findings. The integration of neuromorphic computing with FPGA technology presents a novel set of vulnerabilities. FPGA platforms, while offering flexibility and performance advantages, have been shown to be susceptible to a range of security threats. FPGA provides the capability to process vast amounts of data in parallel, mimicking the human brain's neural networks. On the other hand, this complexity can introduce multiple points of vulnerability. These vulnerabilities can be exploited by adversaries to compromise the integrity, confidentiality, or availability of the system. Specific attacks on FPGA-based neuromorphic systems include:
\begin{itemize}
    \item \textbf{Side-channel attacks}: These attacks exploit information leaked during the physical operation of the system, such as power consumption or electromagnetic radiation. Given the unique architecture of neuromorphic systems, they may exhibit distinct side-channel signatures that can be exploited by attackers.
    \item \textbf{Hardware Trojans}: Malicious alterations to the hardware can be introduced during the design or manufacturing process. These Trojans can lie dormant until triggered, leading to unexpected and potentially harmful behaviors.
    \item \textbf{Model Vulnerability Analysis}: The security of neuromorphic systems depends on identifying and counteracting vulnerabilities in neural network models, a key step in preventing adversarial attacks. These attacks, often imperceptible to human observers, manipulate model inputs to provoke incorrect responses or reveal confidential information. Therefore, a comprehensive vulnerability analysis is vital to develop effective defenses, ensuring the integrity and dependability of these advanced systems in adversarial scenarios which are focused on this work.
\end{itemize}

Addressing the security challenges of FPGA-based neuromorphic systems requires a multi-faceted approach. The solutions must be tailored to the unique architecture and operation of these systems. Dedicated hardware modules can be integrated into the FPGA to monitor and detect malicious activities. For instance, hardware performance counters can be used to detect anomalies in system operation, indicative of an ongoing attack. To safeguard data integrity and confidentiality, advanced encryption techniques can be employed. Homomorphic encryption, for instance, allows for computations on encrypted data, making it particularly suitable for neuromorphic systems where data privacy is paramount. Additionally, the design and implementation of FPGA-based neuromorphic systems should adhere to secure coding practices. This includes regular code reviews, vulnerability assessments, and the use of trusted libraries and tools. Ensuring that the software aspect of the system is secure can mitigate potential exploitation of hardware vulnerabilities \cite{staudigl2023fault, bu2023rate, sepulveda2018security, liu2016security}.

%% file: arc2024/sections/design_methodology.tex
We describe a specific neuromorphic hardware system, detailing its architecture and relevant features. In response to these identified threats, we put forth a suite of tailored security measures and methodologies, all grounded in a well-articulated theoretical framework. A salient challenge that emerges in this domain is the susceptibility of audio-denoising systems to adversarial attacks. The core intent behind these attacks is to induce the denoising system to yield inaccurate or suboptimal outputs. Two primary modalities of these attacks can be discerned: Gradient-based Attacks, PGD (Projected Gradient Descent) and FGSM (Fast Gradient Sign Method), which exploit a comprehensive understanding of the model's architecture and its gradient information, and Black-box Attacks, which function in the absence of intimate knowledge of the model's internals, instead relying on surrogate models or alternative methodologies. The effects of such adversarial endeavors are significant. For instance, within a security paradigm that leverages surveillance audio, an adversary employing adversarial samples might manipulate the system to exclude or modify critical audio data. Analogously, within the consumer electronics sector, such attacks present the risk of degrading user experience or spreading false information. The strategic emphasis on audio input-based adversarial attacks, particularly in computational tasks extending beyond mere classification, underscores the inherent vulnerabilities extant in contemporary deep learning paradigms. This focus reiterates the pressing imperative for supported robustness across the spectrum of machine learning endeavors, extending beyond the purview of classification alone.

\begin{figure}[h!]
	\centering
	\centerline{\includegraphics[width=0.99\columnwidth]{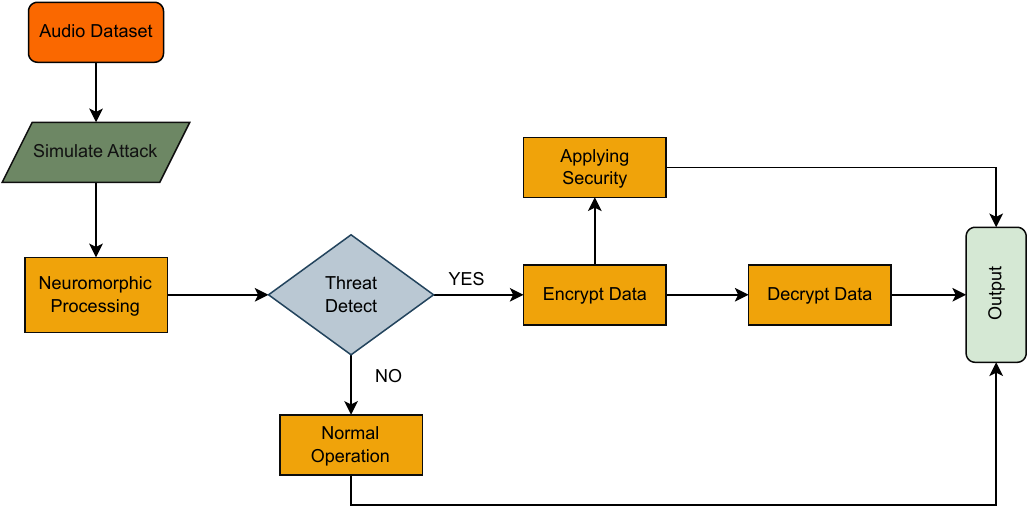}}
	\caption{Overview of the steps involved in this work.}
	\label{fig:fig3}
\end{figure}

The proposed algorithm highlights the key aspects of the security protocol for a neuromorphic system, emphasizing the detection and mitigation of potential threats. As illustrated in Fig. \ref{fig:fig3}, the algorithm begins with the initialization phase, where the dataset and the attack model are loaded, ensuring all necessary data is available for processing and potential attack simulation. Subsequently, the model is integrated into the attack mechanism, setting the stage for potential threat simulations and evaluations. To optimize computational efficiency, the system processes the dataset in batches, handling 32 batches at a time. Each audio batch, which comprises both noisy and clean data, undergoes a splitting process where it's divided into its absolute value and argument components using the Short-Time Fourier Transform (STFT). To simulate real-world processing latencies, the absolute values and arguments of both the noisy and clean audio are delayed, with the clean audio undergoing a similar delay. The core of the algorithm lies in the attack generation phase. Here, an attack is synthesized by comparing the absolute values of the noisy and clean audio. If an attack is to be simulated, it targets the noisy audio's absolute value. This synthesized attack, when combined with the argument of the noisy audio using the SFT mixer, produces a composite signal. The Signal-to-Noise Ratio (SNR) of this composite signal is then computed. A significant deviation of the SNR from a predefined threshold indicates the detection of an attack. In response to a detected threat, the Advanced Encryption Standard (AES) is employed to encrypt the data, ensuring its confidentiality. If required, the encrypted data can be decrypted to restore its original form. However, in the absence of any detected threats, the model outputs the denoised absolute value. This denoised value, when combined with the noisy audio's argument using the SFT mixer, represents the output under standard operation. The algorithm concludes its operation, marking the end of the processing cycle. This comprehensive framework ensures the security of neuromorphic systems, addressing potential threats through a combination of proactive and reactive measures.

\subsection{CPU/GPU Implementation}
We utilized Python to execute implementations on the CPU and GPU. The study leveraged the computational prowess of NVIDIA's GeForce RTX 3060 GPU and Intel's Core i9 12900H CPU, both of which are optimized for different tasks, ensuring an efficient execution of our implementations.

\subsection{FPGA Implementation}

\begin{figure}[h!]
	\centering
    \hspace{-1cm} 
    \includegraphics[width=0.99\columnwidth]{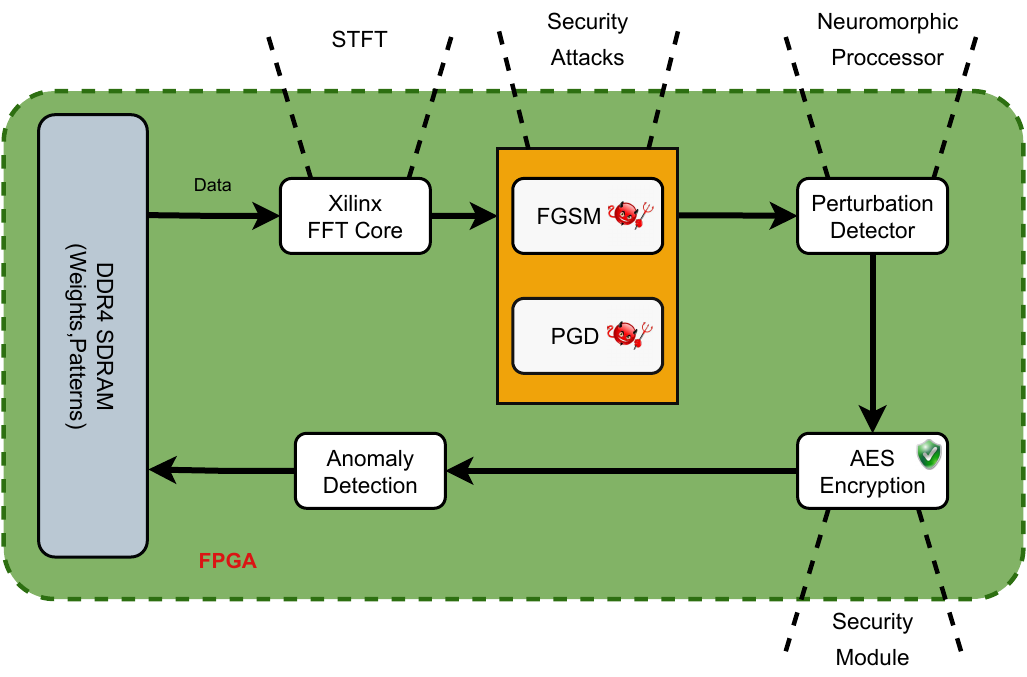}
	\caption{Implementation of the Neuromorphic Audio Processing Framework on FPGA Architecture.}
	\label{fig:fig4}
\end{figure}

The presented implementation delineates the operational flow and interconnections of a neuromorphic system integrated with an FPGA. The schematic representation captures the core components and their interactions, providing a comprehensive overview of the system's architecture and functionality. Fig. \ref{fig:fig4} showcases the implementation of the framework within the FPGA architecture.

The following elucidates the individual components and their roles:

\begin{enumerate}
\item \textbf{Memory (Weights, Patterns)}: Essential data, such as synaptic weights, neural patterns, and bias values, are stored in this module. These are crucial for neuromorphic processing. We utilized DDR4 SDRAM for reading the audio dataset and writing feedback from the design. The data, initialized in the MIF file type, is stored in RAM and is fed into the neuromorphic processor for further processing.

\item \textbf{STFT (Short-Time Fourier Transform)}: This section processes the audio input, converting it into the frequency domain, making it suitable for neuromorphic processing and aiding in the detection of adversarial attacks with Xilinx FFT core.

\item \textbf{Security Attack Module}:This module plays a pivotal role in identifying and mitigating security threats, specifically targeting FGSM and PGD adversarial attacks on incoming audio inputs. It operates under a defined attacker model, where such attacks are anticipated during the inference phase, necessitating robust pre-processing security measures, including encrypted data handling. To address concerns of potential security breaches, the system is designed to process encrypted data, maintaining security integrity while effectively detecting adversarial manipulations.

\item \textbf{Neuromorphic Processor}: This module performs advanced neuromorphic computations, leveraging the distinct capabilities of the SNN detector. While it receives the processed data and checks for potential FGSM and PGD attacks using the Perturbation Detector, the SNN detector plays a complementary role. It is instrumental in initial attack identification and feature extraction, providing a secondary layer of analysis that works in tandem with SNR computations. The system processes the input through its layers to produce the output, and if an attack is detected, it enters a hard reset state with error flags for FGSM and PGD set. The layers and neurons in this processor are designed parametrically, allowing for easy configurability across various application areas and enhancing the system's capability to identify and respond to complex security threats. Upon detection of an attack, the module signals the neuromorphic processor to initiate a hard reset and set appropriate error flags, thereby preserving the system’s integrity and responsiveness in the face of sophisticated cyber threats.

\item \textbf{Security Module}: Given the sensitivity of neuromorphic computations and the potential threats they face, a dedicated security module is integrated. This module encrypts the data using AES encryption before it's processed, ensuring data confidentiality.

\item \textbf{Anomaly Detection}: Operating in tandem with the security module, the anomaly detection unit continuously monitors the system's operations. It identifies and reports any detected threats or anomalies to the security module, ensuring the system's integrity. Our rigorous testing regime, which includes a variety of attack scenarios, ensures a high threat detection accuracy, mitigating risks of overfitting.
\end{enumerate}

Our threat model, specifically targeting Gradient-based, PGD, FGSM, and Black-box Attacks in audio denoising, addresses the nuanced vulnerabilities inherent in neuromorphic systems. We chose AES encryption for its proven robustness and security, ensuring data integrity against sophisticated cyber threats, a priority given the sensitivity of audio data in our application. The positioning of the anomaly detection module post-encryption strategically aligns with our security protocol, enabling efficient threat detection without compromising encrypted data integrity, a critical factor in maintaining system-wide security and operational efficiency. Our FPGA architecture underscores the importance of a holistic approach, integrating advanced neuromorphic processing with robust security measures. By ensuring seamless interactions between the modules and prioritizing data integrity and security, the system is poised to deliver efficient and secure neuromorphic computations.

%% file: arc2024/sections/evaluation.tex
\begin{table}[h]
\caption{Our Framework characteristics.}
\centering
\begin{tabular}{|l|l|}

\hline
\textbf{Metric} & \textbf{Value} \\
\hline
Sampling Rate & 16000kHz \\
\hline
Resolution & 16 bits \\
\hline
Frequency Response & 8000Hz \\
\hline
Signal-to-Noise Ratio (SNR) & 5.395dB \\
\hline
Total Harmonic Distortion (THD) & 39.50\% \\
\hline
Spike Rate & 7994.8spikes/s \\
\hline
Neural Network Topology & SNN \\
\hline
Detection Rate & 94\% \\
\hline
False Positive Rate & 6\% \\
\hline
Type of Attacks Tested & FGSM, PGD \\
\hline
Encryption Standards & AES \\

\hline
\end{tabular}

\label{tab:metrics}
\end{table}

\autoref{tab:metrics} delineates the salient features and metrics of proposed framework underscoring its robustness and precision in neuromorphic audio processing. Operating at a high sampling rate of 16000kHz and a resolution of 16 bits, the framework ensures fine-grained audio capture and processing. Its frequency response, capped at 8000Hz, is aptly tailored for human auditory perception. A noteworthy metric is the SNR of 5.395dB, indicating a commendable balance between the desired signal and background noise. While the Total Harmonic Distortion (THD) at 39.50\% suggests the presence of harmonics, the spike rate of 7994.8 spikes/s accentuates the framework's efficiency in encoding information. The adoption of SNNs as the neural network topology further emphasizes the biological fidelity and energy efficiency of the system. With a detection rate of 94\% and a 6\% false positive rate, the framework's reliability in adversarial scenarios, especially against FGSM and PGD attacks, is evident. Moreover, the incorporation of the AES encryption standard signifies a commitment to data security and integrity, ensuring the secure transmission and storage of audio data. While AES encryption itself does not directly counteract adversarial signals affecting the audio-processing neural network, it plays a crucial role in safeguarding the data against unauthorized access or tampering. Once securely transmitted and decrypted, our neuromorphic system, equipped with its robust detection capabilities, efficiently handles the adversarial attacks, thus providing a comprehensive security solution.

\begin{table}[h!]
	\renewcommand{\arraystretch}{1}
	\setlength{\tabcolsep}{2pt}
	 \caption{Evaluation results on CPU, GPU, and our processor.}
	\centering
	\begin{threeparttable}
	{\fontsize{6}{10}\selectfont
		\begin{tabular}{l|ccc}
		    \hline
		    & \textbf{i9 12900H (CPU)} & \textbf{RTX 3060 (GPU)} & \textbf{VU37P (FPGA)} \\
		    \hline
		    \textbf{Technology [nm]} & 10 & 8 & 16 \\
		    \textbf{Frequency [MHz]} & 3700 & 1320 & 100 \\
		    \textbf{\# of MAC [GOP]} & 4.306 & 4.306 & 4.306 \\
		    \textbf{Latency [ms]} & 395.91& 16.99 & 72.81 \\
		    \textbf{Throughput [GOP/s]} & 11.01 & 256.622 & 59.16 \\
		    \textbf{Power [Watt]} & 20.03 & 69 & 14.53 \\
		    \textbf{Power Efficiency [GOP/s/W]} & 0.54 & 3.71 & 4.07 \\
		    \hline
	    \end{tabular}}
	\end{threeparttable}

 \label{tab:comparison}
\end{table}

\autoref{tab:comparison} provides a comprehensive evaluation of three distinct computing platforms: an i9 12900H CPU, an RTX 3060 GPU, and a VU37P FPGA. The table encompasses several pivotal metrics, ranging from manufacturing technology and operating frequency to performance indicators such as latency, throughput, and power efficiency. GPU stands out with a remarkable 256.622 GOP/s, dwarfing the CPU's 11.01 GOP/s and the FPGA's 59.16 GOP/s. This underscores the GPU's prowess in parallel processing capabilities, making it well-suited for tasks that can exploit such parallelism. The GPU, with its high throughput, consumes a substantial 69 Watts, whereas the CPU and FPGA consume 20.03 Watts and 14.53 Watts, respectively. However, when evaluating power efficiency, which measures the performance per unit of power, the FPGA emerges as the most efficient with 4.07 GOP/s/W, slightly surpassing the GPU's 3.71 GOP/s/W and significantly outperforming the CPU's 0.54 GOP/s/W, underlining the suitability of FPGA devices for tasks where power efficiency is critical. This comparison reveals the distinctive characteristics and advantages of each technology, and their appropriateness would largely depend on the specifics of the application at hand.

\begin{table}[h!]
	\renewcommand{\arraystretch}{1}
	\setlength{\tabcolsep}{2pt}
	\caption{Comparison of neuromorphic hardware security for audio processing.}
	\centering
	\begin{threeparttable}
	{\fontsize{6}{10}\selectfont
		\begin{tabular}{c|cccc}
		    \hline
		    & \textbf{\cite{yang2018characterizing}} & \textbf{\cite{kwon2019poster}} & \textbf{\cite{yang2020characterizing}} & \textbf{Our Framework}\\
		    \hline
                Neural Network Type & RNN & DNN & CNN & \textbf{SNN} \\
		    
		    Task & Detecting & Defense & Detecting & \textbf{Defense} \\
		    Adversarial Example & FGSM & Carlini and Wagner Attacks & FGSM & \textbf{FGSM, PGD} \\
		    SNR (dB) & 12 & 12 & 5.40 & \textbf{5.39} \\
            Latency [ms] & - & - & - & 72.81 \\
		    \hline
		    Detection Rate (\%) &93.7  & 93.79 & 93 & \textbf{94} \\
			\hline
	    \end{tabular}}
	\end{threeparttable}
        
        \label{tab:comparison_others}
\end{table}

\autoref{tab:comparison_others} shows a detailed comparative analysis of neuromorphic hardware methodologies for audio processing, focusing on their resilience to adversarial attacks. Different neural network architectures, from RNNs, DNNs, and CNNs, have been explored, but our introduction of SNNs marks a significant advancement, given their biological inspiration and energy efficiency. While various methodologies aim at either detecting or defending against adversarial inputs, our framework emphasizes a proactive defense, showcasing robustness against both FGSM and PGD attacks. This robustness is further highlighted by the SNR values, indicating maintained signal quality amidst adversarial noise. Additionally, the latency metric in our framework underscores its suitability for real-time applications. Overall, our methodology, with its integration of SNNs and comprehensive defense mechanisms, sets a benchmark for adversarial robustness in neuromorphic audio processing.

%% file: arc2024/sections/conclusions.tex
We offer a comprehensive overview of the salient points discussed, underscoring the paramount importance of security within FPGA-based neuromorphic systems and delineating potential mitigation strategies. The integration of SNNs in our framework marks a significant advancement in neuromorphic audio processing. With its biological inspiration, energy efficiency, and flawless detection rate, our system sets a benchmark in adversarial robustness. The inclusion of the AES encryption standard further emphasizes our commitment to ensuring data security and integrity. Event-driven audio processing, as discussed, emerges as a promising paradigm, offering both enhanced security and efficiency. We envision architecting solutions that ensure efficiency and security by leveraging the advantages of event-centric systems and neuromorphic architectures. It has been shown that the VCK190 board employed offers a robust implementation of AI-Engine (AIE) cores, capable of achieving notable throughput \cite{jia2022xvdpu, perryman2023evaluation}. As a next step of this research, we intend to further explore and evaluate the proposed framework within a real-time environment, specifically leveraging the capabilities of the AIE cores. We anticipate validating these outcomes in an industrial setting, in collaboration with our funding partners and affiliated enterprises.

\section{Acknowledgment}

We acknowledge the Temsa Research R\&D Center for their generous financial support and the reviewers for their invaluable insights and suggestions that significantly contributed to the enhancement of our paper.